\documentclass[preprint, showpacs,preprintnumbers,amsmath,amssymb,nofootinbib]{revtex4}
\usepackage{amssymb}
\usepackage{amsfonts}
\usepackage{CJK}

 \usepackage{epsf}
\usepackage{graphicx}  
\usepackage{dcolumn}   
\usepackage{bm}        
\usepackage{subfigure}
\usepackage{float}
\begin{document}
\newcommand{\be}[1]{\begin{equation}\label{#1}}
 \newcommand{\ee}{\end{equation}}
 \newcommand{\bea}{\begin{eqnarray}}
 \newcommand{\eea}{\end{eqnarray}}
 \newcommand{\bed}{\begin{displaymath}}
 \newcommand{\eed}{\end{displaymath}}
 \def\disp{\displaystyle}

\def\etal{et al.~}
\def\aa{Astron. \& Astrophys.}
\def\aj{Astron. J.~}
\def\apj{Astrophys. J.~}
\def\apjl{Astrophys. J. Lett.~}
\def\araa{Ann. Rev. Astron. Astrophys.~}
\def\apjs{Astrophys. J. Suppl. Ser.~}
\def\cqg{Class.~Quant.~Grav.~}
\def\jcap{J.~Cosmol.~Astropart.~Phys.~}
\def\jhep{J.~High~Energy~Phys.~}
\def\pasp{Publ. Astron. Soc. Pac.~}
\def\plb{Phys. Lett. B~}
\def\prl{Phys. Rev. Lett.~}
\def\prd{Phys. Rev. D~}
\def\raa{Res.~Astron.~Astrophys.~}
\def\mnras{Mon. Not. R. Astron. Soc.~}

\def\gsim{ \lower .75ex \hbox{$\sim$} \llap{\raise .27ex \hbox{$>$}} }

\def\lsim{ \lower .75ex \hbox{$\sim$} \llap{\raise .27ex \hbox{$<$}} }

\title{Constraints on the exponential $f(R)$ model from latest Hubble parameter measurements}

%
\author{Yun Chen$^{1,2,3}$, Chao-Qiang Geng$^{1,2}$, Chung-Chi Lee$^2$, Ling-Wei Luo$^1$, and Zong-Hong Zhu$^{4}$}

\address{$^1$Department of Physics, National Tsing Hua University, Hsinchu 300, Taiwan
\\$^2$Physics Division, National Center for Theoretical Sciences, Hsinchu 300, Taiwan
\\$^3$Key Laboratory of Computational Astrophysics, National Astronomical Observatories, Chinese Academy of Sciences, Beijing, 100012, China
\\$^4$Department of Astronomy, Beijing Normal University, Beijing 100875, China }

\begin{abstract}
We investigate the viable exponential $f(R)$ gravity in the metric formalism with $f(R)=-\beta R_s (1-e^{-R/R_s})$.
The latest sample of the Hubble parameter measurements with 23 data points is used to place bounds on this $f(R)$ model.
A joint analysis is also performed with the luminosity distances of Type Ia supernovae and baryon acoustic oscillations
in the clustering of galaxies, and the shift parameters from the cosmic microwave background measurements, which leads to
$0.240<\Omega_m^0<0.296$ and $\beta>1.47$ at 1$\sigma$ confidence level.
 The evolutions of the deceleration parameter $q(z)$ and the effective equations of state $\omega_{de}^{eff}(z)$
 and $\omega_{tot}^{eff}(z)$ are displayed. By taking the best-fit parameters as prior values, we work out the
 transition redshift (deceleration/acceleration) $z_T$ to be about 0.77. It turns out that the recent observations
 are still unable to distinguish the background dynamics in the $\Lambda$CDM and  exponential $f(R)$ models.

\end{abstract}


 \maketitle
 \renewcommand{\baselinestretch}{1.5}

\section{Introduction}
\label{Intro}

It is one of the most significant problems in cosmology to understand the physical mechanism behind the
late-time acceleration of the Universe~\cite{Riess1998,Permutter1999}. A number of scenarios have
been proposed to account for this phenomenon~\cite{Copeland2006, LiM2011, Mortonson2014}.
In general, these fall into two categories: (i) the existence of an exotic form of energy with negative pressure,
 dubbed as ``dark energy'',  corresponding to a modification of the
 energy-momentum tensor in Einstein equations; and (ii) a modification of gravity,
 such as $f(R)$ gravity models with $f(R)$ representing
 an arbitrary function of the Ricci scalar $R$.
 The $\Lambda$-Cold Dark Matter ($\Lambda$CDM) model is the
 simplest candidate of dark energy and fits a number of observational data well, which is referred to
 as the standard model of Big Bang cosmology. However, this model cannot explain the origin of the
 inflation~\cite{Covi2003, Ratra&Vogeley2008} or the nature of dark energy by itself, and is also embarrassed
  by the well known cosmological constant problems~\cite{Weinberg1989, Carroll2001}, known
   as ``coincidence'' and ``fine-tuning'' problems. The Lagrangian density for $\Lambda$CDM is
   given by $f(R)=-2\Lambda$, where  $\Lambda$ is the cosmological constant.

At present, the cosmological observations cannot distinguish between  dark energy and modified gravity models.
 An important reason for the interest on the modified gravity theories is that the late-time acceleration of the Universe
 can be realized without recourse to an explicit dark energy matter component~\cite{Nojiri2006, Tsujikawa2010, Clifton2012},
 while   $f(R)$ gravity  is one of the popular and simplest modifications to general relativity (GR)~\cite{Sotiriou2010, DeFelice2010, Nojiri2011}.
 There are two approaches to derive field
 equations from the action in  $f(R)$ gravity, i.e., the metric and Palatini formalisms. In GR, the two approaches provide
  identical field equations. However, they give rise to different field equations for the $f(R)$ models with non-linear
  forms of the Lagrangian density. It is pointed out that the Palatini $f(R)$ gravity appears to be in conflict with
  the Standard Model of particle physics~\cite{Sotiriou2010}. Given this, the metric $f(R)$ gravity is preferred.
  The viability of $f(R)$ gravity is examined with various criteria~\cite{Starobinsky2007}, such as the local gravity
  constraints, the presence of the matter-dominated epoch, the stability of cosmological perturbations, the stability
  of the late-time de Sitter point, and avoiding anti-gravity.

  The exponential $f(R)$ gravity theory in the
  metric formalism is one of the viable models, which contains only one
   more parameter than the $\Lambda$CDM model and has been broadly studied in the
   literature~\cite{Linder2009,Zhang2006c,Zhang2007,Tsujikawa2008,Cognola2008a,Ali2010,
   Bamba2010a, Bamba2010b, Yang2010, Yang2011, Elizalde2011,Geng:2012zc,Lee2012,Bamba:2012qi}.
   In this work, we investigate the constraints on this model from the latest  measurement on the Hubble parameter $H(z)$.
   In addition, several other popular probes are also associated in the joint analysis, including the
   distance measurements of baryon acoustic oscillations (BAO) from 6dFGS, WiggleZ and SDSS III Data
   Release 11 (DR11) and  Type Ia supernovae (SNe Ia) from the SCP Union2 compilation,
   as well as the cosmic microwave background (CMB) measurements of the shift parameters
  from Planck and WMAP-9. The dynamical features of the model are also analyzed,
  such as the evolutions of the effective equations of state
  (EoS) $\omega_{de}^{eff}$ and $\omega_{tot}^{eff}$, and the deceleration parameter $q(z)$.
  Note that  the data on $H(z)$ were not employed in the previous study of the exponential  gravity in Ref.~\cite{Yang2010}.

The paper is organized as follows. In Sec.~\ref{ModelReview}, we review the exponential $f(R)$ gravity model.
In Sec.~\ref{ObsConstr}, we examine constraints on the model from different observational samples,
particularly the latest $H(z)$ sample, along with the recent BAO, SNe Ia and CMB data sets.
In Sec.~\ref{analysis}, we analyze the results of observational constraints and discuss several noticeable problems.
Finally, we present our main conclusions in Sec.~\ref{summary}.

\section{Exponential $f(R)$ gravity in the metric formalism }
\label{ModelReview}

In  $f(R)$ gravity, the 4-dimensional action is given by
\begin{equation}
\label{eq:fR_action}
S=\frac{1}{2\kappa^2}\int d^4x \sqrt{-g} [R+f(R)]+S_m,
\end{equation}
where $\kappa^2=8\pi G$,  $f(R)$ is a general function of the Ricci scalar $R$, and $S_m$ is the matter action.
In this work, we study on an exponential $f(R)$ model in the metric formalism~\cite{Linder2009,Bamba2010a} with
\begin{equation}
\label{eq:fR_form}
f(R)=-\beta R_s(1-e^{-R/R_s}),
\end{equation}
where $R_s$ is a characteristic curvature scale and the combination $\beta R_s$ can be determined by the present
matter density $\Omega_m^0$. We take $(\Omega_m^0, \beta)$ as the free-parameter pair in this model.
As pointed out in Ref.~\cite{Linder2009}, there is no valid attractor solution for $\beta<1$ because this would require $R<0$.
Given this, the prior $\beta \geq 1$ is taken in our calculation.
Throughout this paper, a subscript 0 denotes the evaluation at the present time.

Based on the action given by Eq.~(\ref{eq:fR_action}), one can obtain the modified Friedmann equation of motion~\cite{Song2007}
\begin{equation}
\label{eq:FriedmannEq}
H^2=\frac{\kappa^2 \rho_m }{3}+\frac{1}{6}(f_RR-f)-H^2(f_R+f_{RR}R'),
\end{equation}
where $H\equiv \dot{a}/a$ is the Hubble parameter, $a$ is the cosmic scale factor, and
$\rho_m$ is the energy density of matter including both cold dark and baryonic matters.
In this study, we use a dot for the derivative with respect to the cosmic time $t$, a prime for $d/d\ln a$,
$f_R\equiv\partial{f}/\partial{R}$, and $f_{RR}\equiv \partial^2{f}/\partial{R^2}$.
We consider the flat FLRW spacetime
with the metric  $(-, +, +, +)$, in which the Ricci scalar is given by
\begin{equation}
\label{eq:R_Eq}
R=12H^2+3(H^2)'.
\end{equation}

Following Refs.~\cite{Yang2010,Hu2007}, we define
\begin{equation}
y_{H}\equiv \frac{H^{2}}{m^{2}}-a^{-3},\quad y_{R}\equiv\frac{R}{m^{2}}-3a^{-3},
\label{eq:yH_yR}
\end{equation}
where $m^2\equiv \kappa^2 \rho_m^0$ with $\rho_m^0$  the present matter density.
With Eqs.~(\ref{eq:FriedmannEq}) and (\ref{eq:yH_yR}), one can get a second order differential equation of $y_H$,
\begin{equation}
y_{H}''+J_{1}y_{H}'+J_{2}y_{H}+J_{3}=0,
\label{eq:yH_2order}
\end{equation}
 where
 \begin{eqnarray}
J_{1}&=&4-\frac{1}{y_{H}+a^{-3}}\frac{f_{R}}{6m^{2}f_{RR}},\nonumber\\
J_{2}&=&-\frac{1}{y_{H}+a^{-3}}\frac{f_{R}-1}{3m^{2}f_{RR}},\nonumber\\
J_{3}&=&-3a^{-3}+\frac{f_{R}a^{-3}+f/3m^{2}}{y_{H}+a^{-3}}\frac{1}{6m^{2}f_{RR}},
\end{eqnarray}
 with
 \begin{eqnarray}
R=m^{2}\left[3\left(y_{H}'+4y_{H}\right)+3a^{-3}\right].
\end{eqnarray}
The evolution of the Hubble parameter can be obtained by solving Eq.~(\ref{eq:yH_2order}) numerically.
We note that it is very difficult to find the true solutions of Eq.~(\ref{eq:yH_2order}) when the exponential $f(R)$ model
behaves essentially like the $\Lambda$CDM model (i.e., $e^{-R/R_s}<10^{-5}$) in the high redshift regime.
As a result, the evolution of the Hubble parameter in this high-z regime can be computed with the following equation,
\begin{equation}
H(z)= H_{0}\sqrt{\Omega_{m}^{0}\left(1+z\right)^{3}+\Omega_{r}^{0}\left(1+z\right)^{4}+\frac{\beta R_{S}}{6H_{0}^{2}}},
\label{eq:FriedmannEq_LCDM}
\end{equation}
where $\Omega_{\Lambda}=\beta R_s/6H_0^2\cong \Omega_m^0 y_H(z_{high})$ is the dark energy density, and $\Omega_r^0$ is
 the energy density parameter of radiation which should not be neglected.
 In the latter analysis, we employ the prior, i.e., $\Omega_r^0/\Omega_m^0=1/(1+z_{eq})\approx 2.9\times 10^{-4}$, based on the Planck 2013 results~\cite{Planck2013}.

Treating the modified terms in  the Friedmann equation of Eq.~(\ref{eq:FriedmannEq}) as an effective dark energy density,
one can define $\rho_{de}^{eff}\equiv3H^2/\kappa^2-\rho_m$
 and $\Omega_{de}^{eff}\equiv\rho_{de}^{eff}/(3H^2/\kappa^2)=y_H/(y_H+a^{-3})$.
Furthermore, we write the effective dark energy EoS, $\omega_{de}^{eff}$, based on the effective conservation equation,
 $\dot{\rho}_{de}^{eff}+3H(1+\omega_{de}^{eff}){\rho}_{de}^{eff}=0$,
to be
\begin{eqnarray}
w_{de}^{eff}=-1-\frac{y_{H}'}{3y_{H}}.
\label{eq:w_de}
\end{eqnarray}
Analogously, one can also define the total effective EoS, $\omega_{tot}^{eff}$ in terms of
 $\dot{\rho}_{tot}^{eff}+3H(1+\omega_{tot}^{eff}){\rho}_{tot}^{eff}=0$ and $\rho_{tot}^{eff}=\rho_m + {\rho}_{de}^{eff}$, i.e.,
 \begin{eqnarray}
w_{tot}^{eff}=-1-\frac{y_{H}'-3a^{-3}}{3(y_{H}+a^{-3})}.
\label{eq:w_tot}
\end{eqnarray}
According to the definitions of the deceleration parameter $q\equiv -(\ddot{a}a)/\dot{a}^2$ and the Hubble parameter $H\equiv\dot{a}/a$, one obtains
\begin{eqnarray}
q(z)=\frac{1}{2}-\frac{y'_H+3y_H}{2(y_H+a^{-3})}.
\label{eq:q_z}
\end{eqnarray}

\section{Constraints from recent observational data sets}
\label{ObsConstr}
In this section, we place limits on the $\textbf{p}=(\Omega_m^0, \beta)$ parameter space with the recent observational data sets
to update the results in Ref.~\cite{Yang2010}, in which the latest sample of the $H(z)$ measurements was not  employed.
To break possible degeneracies in the $\Omega_m^0$ -- $\beta$ plane, we also perform a joint analysis
involving the BAO measurements from SDSS III, 6dFGS and WiggleZ surveys, the Union2 SNe Ia sample, and the CMB shift parameters
 from the Planck and  WMAP-9. In our analysis, we add a Gaussian prior on the Hubble constant,
  $H_0 = 73.8\pm 2.4 {\rm~km~s^{-1}~Mpc^{-1}}$, from the Hubble Space Telescope (HST) observations~\cite{Riess2011}.

\subsection{The latest Hubble parameter measurements}
\begin{figure}[htbp]
\centering
\includegraphics[width=0.496\linewidth]{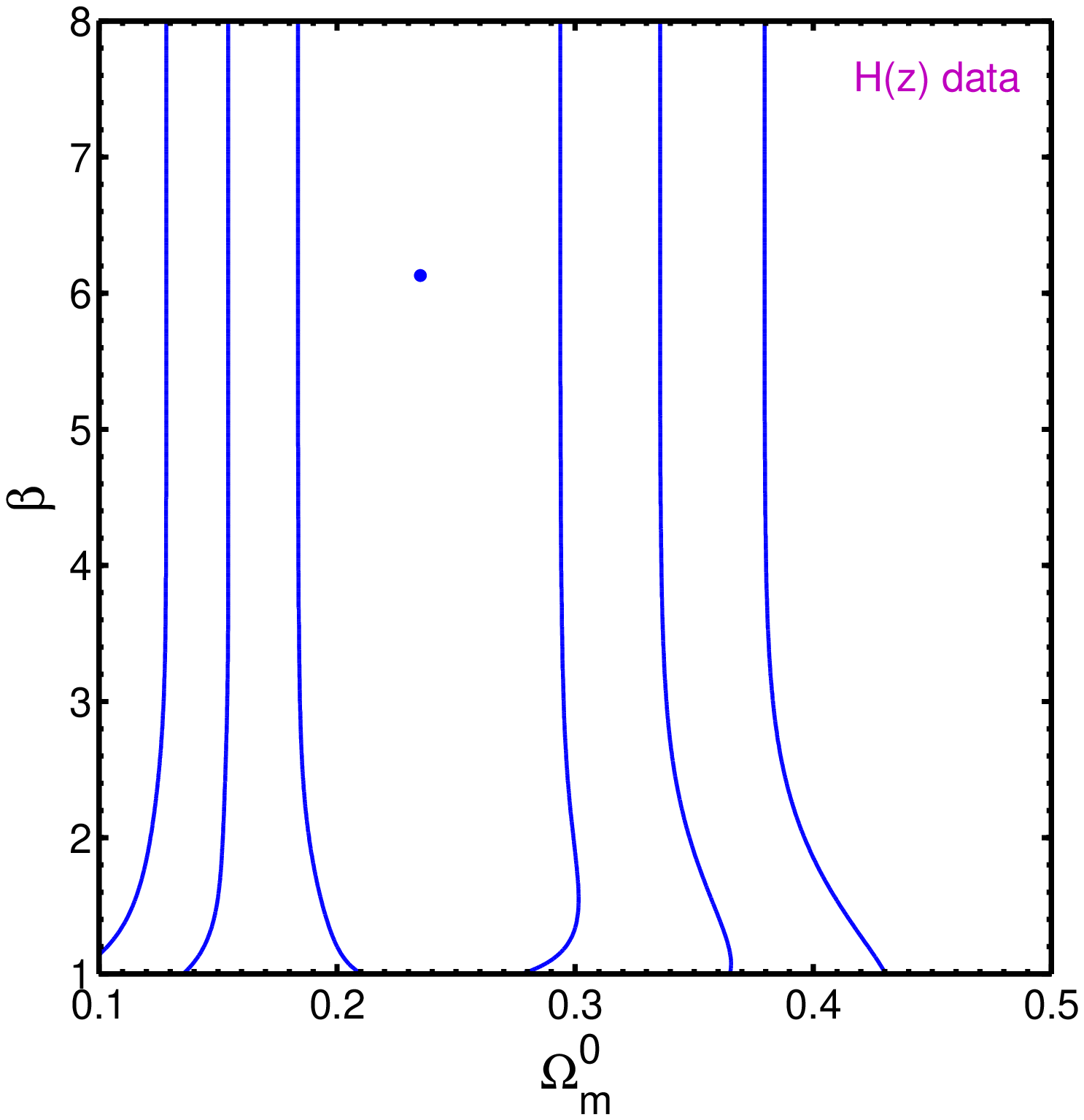}
\includegraphics[width=0.496\linewidth]{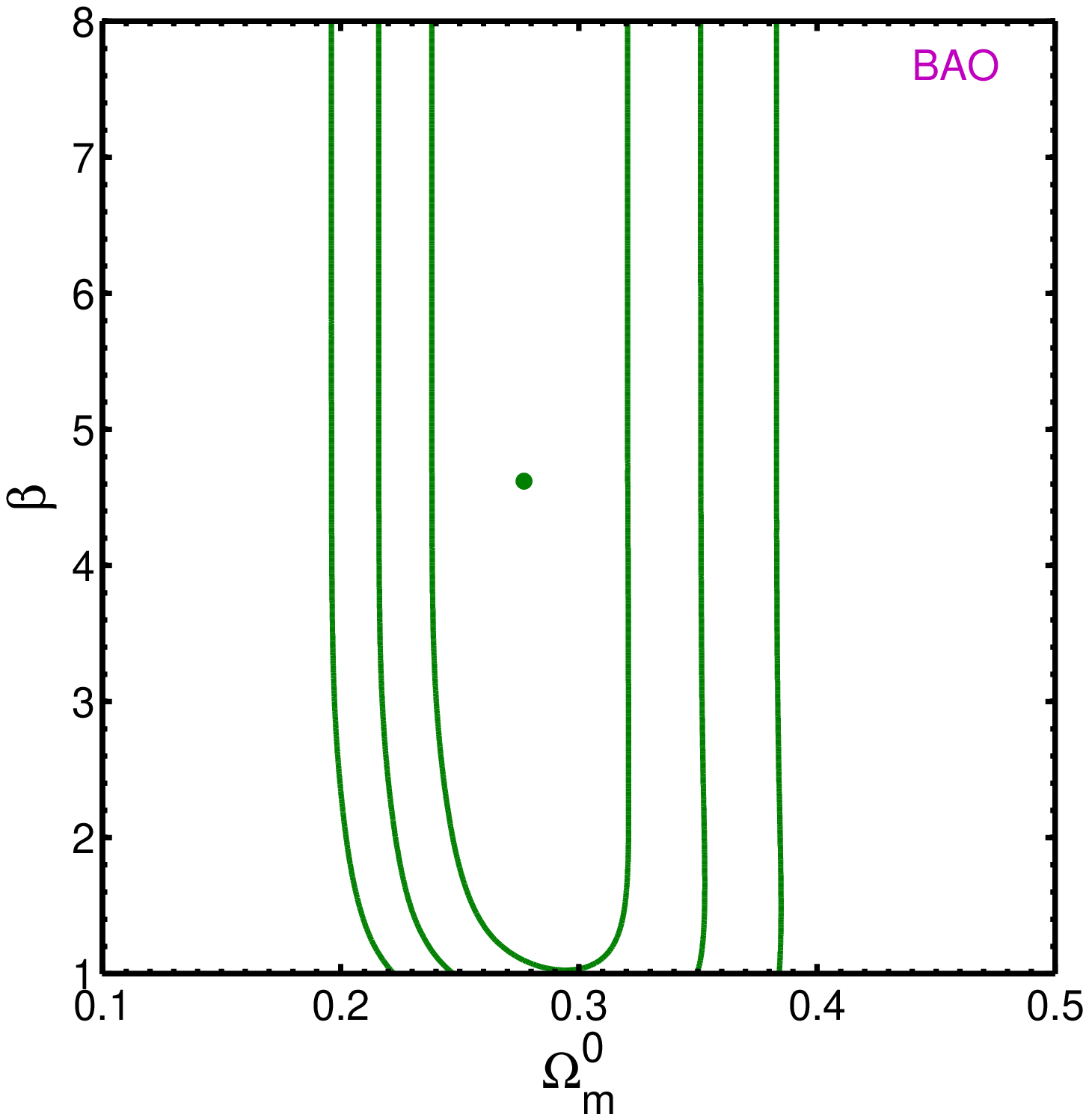} \\
\includegraphics[width=0.496\linewidth]{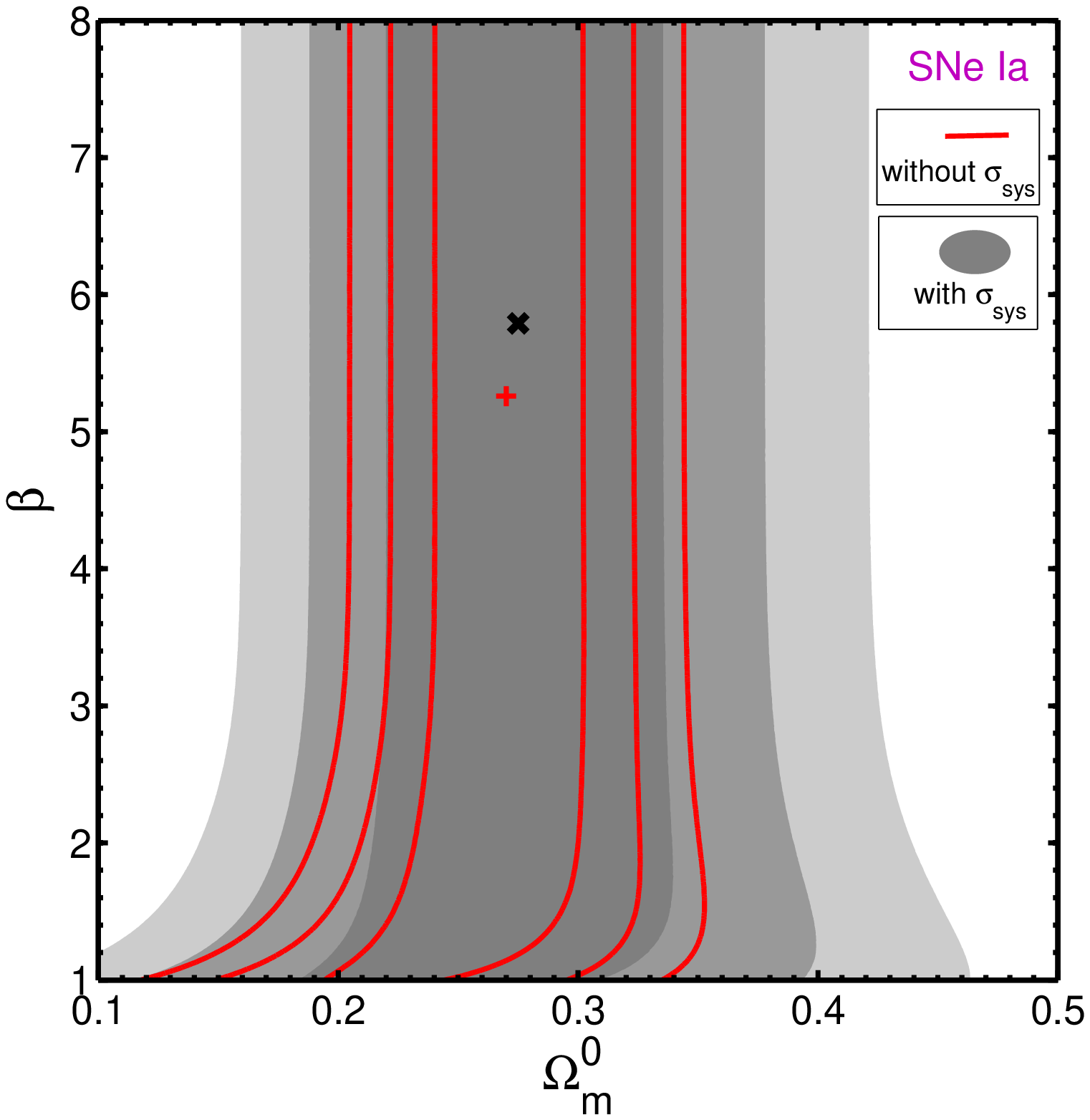}
\includegraphics[width=0.496\linewidth]{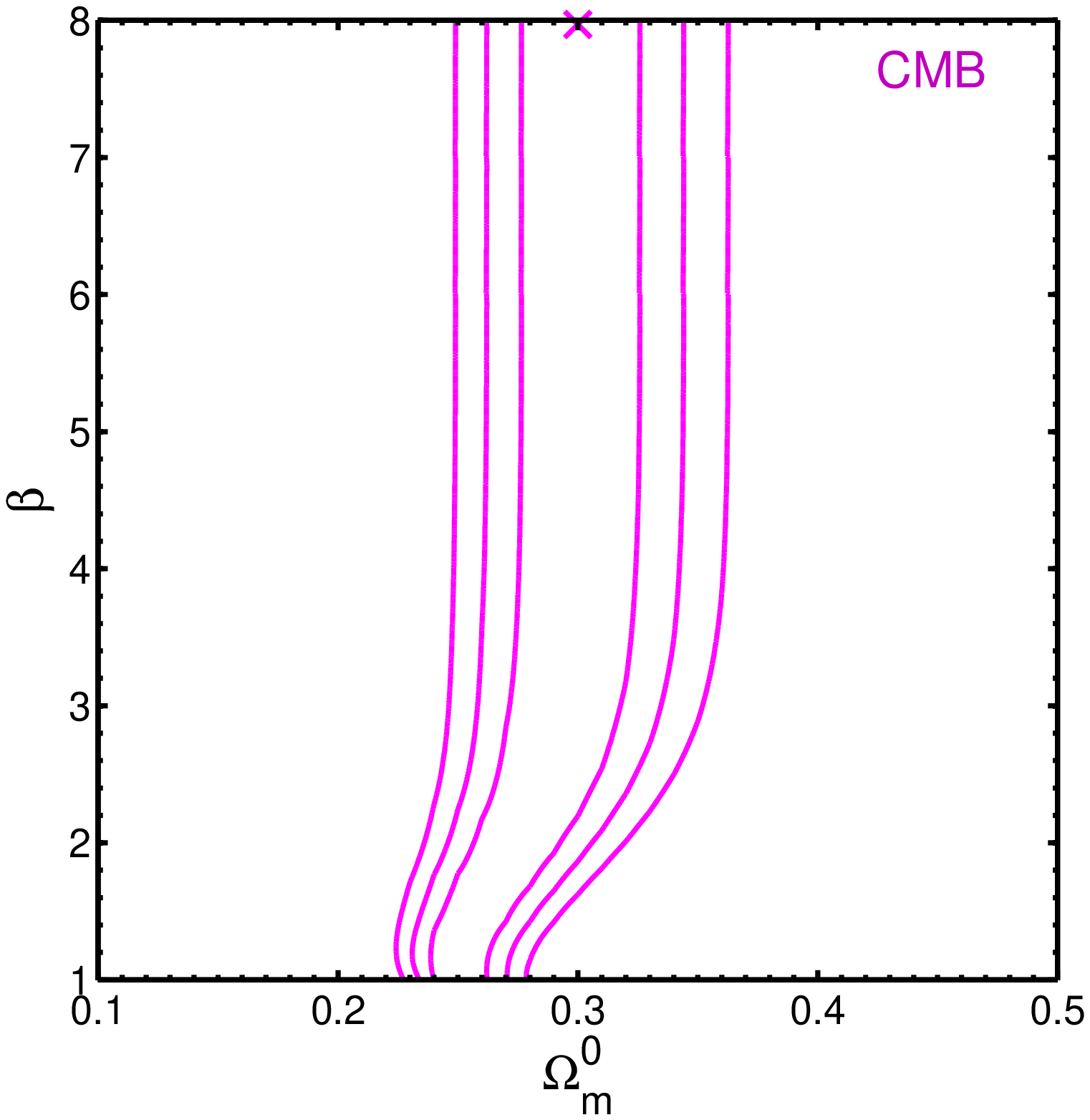}
\caption{Contours correspond to 1, 2 and 3$\sigma$
confidence levels in the ($\Omega_m^0$, $\beta$) plane constrained from
the  $H(z)$ data (\textbf{Upper-left});  BAO measurements (\textbf{Upper-right});
 SNe Ia data with and without systematic errors (\textbf{Lower-left}); and CMB shift parameters (\textbf{Lower-right}).}
\label{fig:HzBAOSNeCMB}
\end{figure}

The recent Hubble parameter measurements are proved to be very powerful in constraining the cosmological
parameters~\cite{Farooq2013, ChenY2013}. The latest $H(z)$ sample listed in Ref.~\cite{ChenY2013}
includes 23 measurements obtained with the differential age (``DA'' for short)~\cite{JimenezLoeb2002, Simon2005, Stern2010, Moresco2012, Zhang2012} method
and 7 ones with the clustering of galaxies or quasars~\cite{Gaztanaga2009, Blake2012, Chuang2013, Anderson2013a, Busca2013}. Since the $H(z)$ measurements from
the clustering are correlated with the BAO data introduced in Sec.~\ref{OtherData}, to keep the independence
we just employ the 23 $H(z)$ measurements with the DA method in our analysis.  The sample contains measurements of
$H_{obs}(z_i)$ at redshifts $z_i$ with the corresponding one-standard deviations $\sigma_{H,i}$. Utilizing these $H(z)$ data,
it is straightforward to put constraints on the model parameters by calculating the corresponding $\chi^2$, given by
\begin{equation}
\chi^2_H(\textbf{p},H_0)=\sum_{i=1}^{23}\frac{[H_{th,i}(\textbf{p},H_0;z_i)-H_{obs}(z_i)]}{\sigma^2_{H,i}},
\label{eq:chi2_Hz}
\end{equation}
where $H_{th,i}(\textbf{p},H_0;z_i)$ can be computed with Eq.~(\ref{eq:yH_2order}). By marginalizing over the nuisance parameter
$H_0$ with the HST prior $H_0 = 73.8\pm 2.4 {\rm~km~s^{-1}~Mpc^{-1}}$, we get the modified $\chi^2_H(\textbf{p})$
with the methodology described in Ref.~\cite{ChenY2013}. The upper-left panel of Fig.~\ref{fig:HzBAOSNeCMB} shows the constraints from the $H(z)$ data at 1, 2 and 3$\sigma$ confidence levels,
corresponding to the sets of cosmological parameters (centered
on the best-fit parameter values $\textbf{p*}$) bounded by $\chi^2_H(\textbf{p})=\chi^2_H(\textbf{p*})+2.3$, $\chi^2_H(\textbf{p})=\chi^2_H(\textbf{p*})+6.17$, and $\chi^2_H(\textbf{p})=\chi^2_H(\textbf{p*})+11.8$, respectively.

\subsection{Other data combinations}
\label{OtherData}
Besides the Hubble parameter measurements, we also consider the following cosmological probes: (i) the measurements of  BAO in the galaxy distribution; (ii) the distance moduli of SNe Ia; and (iii) the CMB shift parameters.

The baryon acoustic oscillations in the primordial plasma have striking effects on the anisotropies of CMB and  the large scale
 structure (LSS) of matter. The distance-redshift measurements of BAO are distilled from the power spectrum of galaxies and calibrated by
 the CMB anisotropy data. It is common to report the BAO distance constraints as a combination of the angular
  diameter distance, $D_A(z)$, and the Hubble parameter, $H(z)$, i.e.,
\begin{equation}
\label{eq:A_bao}
A(z) \equiv \frac{H_0 \sqrt{\Omega_{m}^0} D_V(z)}{c z},
\end{equation}
or
\begin{equation}
\label{eq:dz_bao}
d_z \equiv r_s(z_d)/D_V(z),
\end{equation}
where $D_V(z)$ is the volume-averaged distance defined as $D_V(z)\equiv [(1+z)^2 D_A^2(z)cz/H(z)]^{1/3}$ \cite{Eisenstein2005},
and $r_s(z_d)$ is the comoving sound horizon at the drag epoch $z_d$. For details on the definition and estimation of $z_d$ and $r_s(z_d)$,
 see Refs.~\cite{Yang2010,Eisenstein1998}. The BAO distance measurements employed in this paper are listed below:
  (i) $d_z(z=0.106)=0.336 \pm 0.015$ from the 6dFGS reported by Beutler et al. in Ref.~\cite{Beutler2011};
  (ii) three correlated measurements of $A(z=0.44)=0.474\pm 0.034$, $A(z=0.6)=0.442\pm 0.020$ and $A(z=0.73)= 0.424\pm 0.021$
  from the WiggleZ survey with the inverse covariance matrix listed in Table~2 of Blake et al. in Ref.~\cite{Blake2011};
 and  (iii) $d_z(z=0.32)=0.1181\pm 0.0023$ from LOWZ and $d_z(z=0.57)=0.0726\pm 0.0007$ from CMASS of SDSS-III DR11 galaxy samples reported
   by Anderson et al. in Ref.~\cite{Anderson2013}. The upper-right panel of Fig.~\ref{fig:HzBAOSNeCMB} illustrates the constraints from these
   BAO data at 1, 2 and 3$\sigma$ confidence levels.

The first direct evidence for the cosmic acceleration came from SNe Ia observations~\cite{Riess1998,Permutter1999}.
Here, we use the Supernova Cosmology Project (SCP) Union2 compilation of 557 SNe Ia (covering a redshift range
 $0.015 \leq z \leq 1.4$)~\cite{Amanullah2010}.
Cosmological constraints from the SNe Ia data are obtained by using the distance modulus $\mu (z)$ with its theoretical (predicted) value
\begin{equation}
\label{eq:mu} \mu_{\rm th}(z; \textbf{p}, \mu_0 )=5\log_{10} [D_L(z;
\textbf{p})]+\mu_0,
\end{equation}
where $\mu_0=42.38-5\log_{10}h$ and the Hubble-free luminosity distance is given by
\begin{equation}
\label{eq:DL} D_L(z;\textbf{p})=\frac{H_0}{c}d_L=(1+z)\int_0^z
\frac{dz'}{E(z';\textbf{p})}.
\end{equation}
The best-fit values of cosmological model parameters can be determined by minimizing the $\chi^2$ function
\begin{equation}
\label{eq:chi2_SN1}
\chi^2_{SNe}(\textbf{p}, \mu_0)=\sum_{i,j=1}^{557}[\mu_{th,i}(z_i; \textbf{p}, \mu_0 )-\mu_{obs,i}(z_i)] Cov_{ij}^{-1}[\mu_{th,j}(z_j; \textbf{p}, \mu_0 )-\mu_{obs,j}(z_j)],
\end{equation}
where the nuisance parameter $\mu_0$ can be marginalized over analytically \cite{diPietro2003}.
The modified $\chi^2_{SNe}(\textbf{p})$ is often used in the analysis, i.e.,
\begin{equation}
\label{eq:chi2_SN1}
\chi^2_{SNe}(\textbf{p})=A-\frac{B^2}{C},
\end{equation}
with
 \begin{eqnarray}
A&=&\sum_{i,j=1}^{557}[\mu_{th,i}(z_i; \textbf{p}, \mu_0=0 )-\mu_{obs,i}(z_i)] Cov_{ij}^{-1}[\mu_{th,j}(z_j; \textbf{p}, \mu_0=0 )-\mu_{obs,j}(z_j)],
\nonumber\\
B&=&\sum_{i,j=1}^{557}Cov_{ij}^{-1}[\mu_{th,j}(z_j; \textbf{p}, \mu_0=0 )-\mu_{obs,j}(z_j)],
\nonumber\\
C&=&\sum_{i,j=1}^{557}Cov_{ij}^{-1},
\end{eqnarray}
where $Cov_{ij}^{-1}$ is the inverse of the covariance matrix, which
can be found from the website\footnote{http://supernova.lbl.gov/Union}. The results from the SNe Ia sample
with and without systematic errors are both displayed in the lower-left panel of Fig.~\ref{fig:HzBAOSNeCMB}.

The CMB regarded as the afterglow of the Big Bang can supply us with some  information of the very early Universe.
In addition, the positions of the CMB acoustic peaks contain the information of the cosmic expansion history.
The likelihood of the acoustic scale $l_a$ and the shift parameter $R$ extracted from the CMB angular power spectrum
can allow one to constrain the cosmological parameters~\cite{WangY2007,WangY2008},
where $l_a$ determines the average acoustic
 peak structure and $R$ corresponds to the overall amplitude of the acoustic peaks, given by
 \begin{eqnarray}
l_{A}(z_{*})\equiv(1+z_{*})\frac{\pi D_{A}(z_{*})}{r_{s}(z_{*})},
\end{eqnarray}
and
\begin{eqnarray}
R(z_{*})\equiv\sqrt{\Omega_{m}^{0}}H_{0}(1+z_{*})D_{A}(z_{*}),
\end{eqnarray}
respectively.
Here, $r_s(z_{*})$ is the comoving sound horizon at the photon-decoupling epoch~\cite{Page2003}, and the redshift
to the photon-decoupling surface $z_{*}$ is given by the fitting formula~\cite{Hu1996}
\begin{eqnarray}
z_{*}=1048\left[1+0.00124(\Omega_{b}^{0}h^{2})^{-0.738}\right]\left[1+g_{1}(\Omega_{m}^{0}h^{2})^{g2}\right],
\end{eqnarray}
where
\begin{eqnarray}
g_{1}=\frac{0.0783(\Omega_{b}^{0}h^{2})^{-0.238}}{1+39.5(\Omega_{b}^{0}h^{2})^{0.763}},\quad g_{2}
=\frac{0.560}{1+21.1(\Omega_{b}^{0}h^{2})^{1.81}}.
\end{eqnarray}
Based on  the Planck temperature data combined with the Plank lensing as well as the WMAP polarization at low multipoles ($l\leqslant 23$),
the mean values and the covariance matrix of $(l_a, R, \Omega_b^0 h^2, n_s)$ are obtained in Ref.~\cite{WangY2013}.
In our case, we just utilize the measurements of $(l_a, R, \Omega^0_b h^2)$, where the mean values and covariance matrix are determined with Eqs.~(12), (13) and (16) in Ref.~\cite{WangY2013}, i.e., $l_a(z_{*})=301.37$, $R(z_{*})=1.7407$ and $\Omega_b^0 h^2=0.02228$ with the inverse covariance matrix
\begin{eqnarray}
C_{CMB}^{-1}=\left(\begin{array}{ccc}
43.0180 & -366.7718 & 2972.5\\
-366.7718 & 24873.0 & 4.4650\times10^5\\
2972.5 & 4.4650\times10^5 & 2.1555\times10^7\end{array}\right).
\end{eqnarray}
The CMB data are included by adding the following $\chi^2_{CMB}$,
\begin{eqnarray}
\chi_{CMB}^{2}=\sum_{i=1}^{3}(p_{i}^{th}-p_{i}^{obs})(C_{CMB}^{-1})_{ij}(p_{j}^{th}-p_{j}^{obs}),
\end{eqnarray}
where $p_1=l_a(z_*)$, $p_2=R(z_*)$ and $p_3=\Omega^0_b h^2$. The constraints on the $\textbf{p}=(\Omega_m^0, \beta)$
from the CMB data are presented in the lower-right panel of Fig.~\ref{fig:HzBAOSNeCMB}.

\section{Analysis and discussion}
\label{analysis}

\begin{figure}[htbp]
\centering
\includegraphics[width=0.496\linewidth]{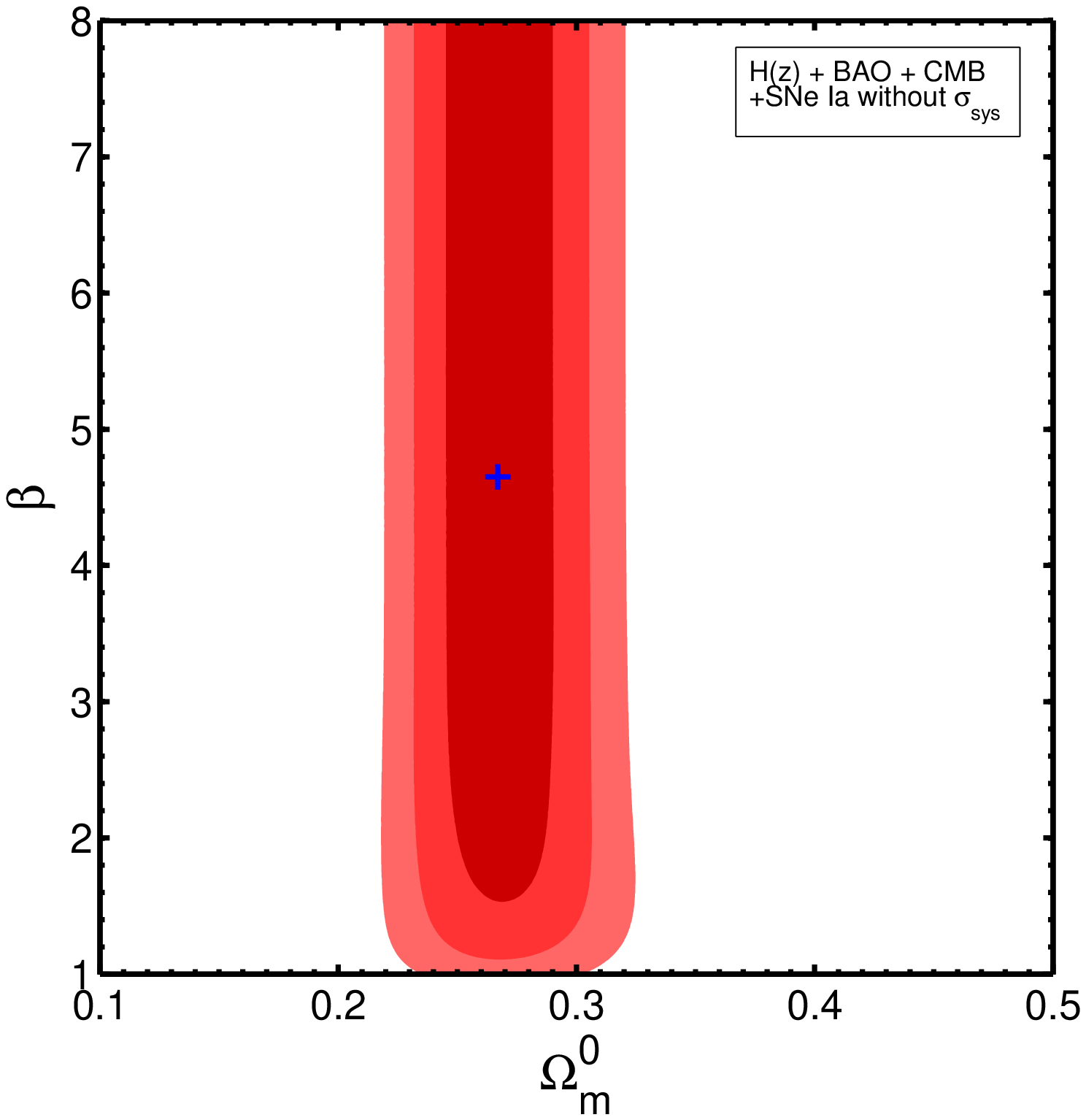}
\includegraphics[width=0.496\linewidth]{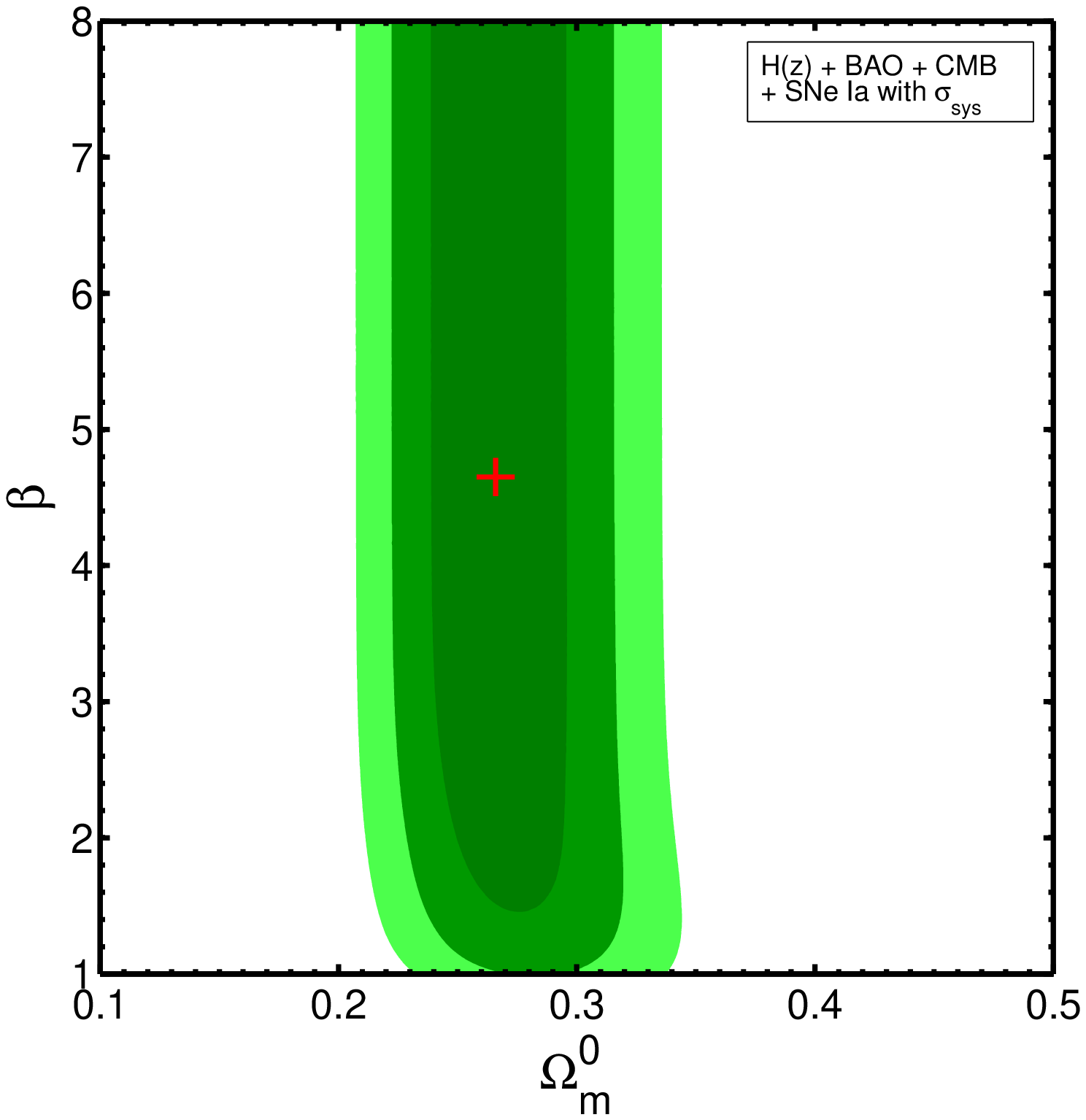}
\caption{Contours correspond to 1, 2 and 3$\sigma$
confidence levels in the ($\Omega_m^0$, $\beta$) plane constrained from joint samples of the CMB and
$H(z)$, BAO, SNe Ia data without (\textbf{left}) and with  (\textbf{right}) systematic errors, respectively.}
\label{fig:JointSamples}
\end{figure}

 From Fig.~\ref{fig:HzBAOSNeCMB}, we see that the current Hubble parameter measurements alone can constrain
the parameters $(\Omega_m^0, \beta)$ significantly. In addition, the constraints from the BAO and SNe Ia measurements
are also restrictive, whereas those from the CMB data are not so tight.
In Fig.~\ref{fig:JointSamples}, we show the joint analyses of the $H(z)$ data with the BAO, CMB and SNe Ia
with (without) systematic errors. Given the complementarity of these data sets, we obtain a considerable enhancement
of the constraining power over $(\Omega_m^0, \beta)$ from the combined fits.
Note that the top sides of the three contours are not closed as we cannot get the upper limit for $\beta$.
For a general illustration, in the joint analysis with
the systematic errors of the SNe Ia data, we work out
$\chi^2_{\beta \rightarrow \infty}=563.58$ with the best-fit value $\Omega_m^0=0.266$.
Obviously, $\Delta \chi^2=\chi^2_{\beta \rightarrow \infty}-\chi^2_{min}=0.01$ is smaller
than $\Delta \chi^2_{1\sigma}(n=2)=2.30$ where $n$ is the number of the parameters
in the model, which implies that $\beta \rightarrow \infty$ is still inside the $1\sigma$ contour.
Basically, the absence of the upper limit for $\beta$ originates from the evolution of the Ricci scalar $R$
with respect to  $z$.
We remark that
$R$ is sensitive (insensitive) to the value of $\beta$ when it is small (large).
Furthermore,
the best-fit value of $\beta$ lies in the range where $R$ is insensitive to  $\beta$, so that
the upper limit of $\beta$  is absent.

The bounces on $(\Omega_m^0, \beta)$ from different data sets are summarized in Table~\ref{tab:results}. 
Comparing the results in Table~\ref{tab:results} of this paper with those in Table 1 of ~\cite{Yang2010},  
we can see that the best-fit values of $\Omega_m^0$ and $\beta$ at 1 $\sigma$ confidence interval are consistent. 
However, it is clear that 
the constraint on the $f(R)$ model  from the $H(z)$ sample is very restrictive. In particular, the  $H(z)$ data alone 
leads to $\beta$ around 6, resulting in that the exponential  gravity is practically undistinguishable from the $\Lambda$CDM.

\begin{figure}[htbp]
\centering
\includegraphics[width=0.496\linewidth]{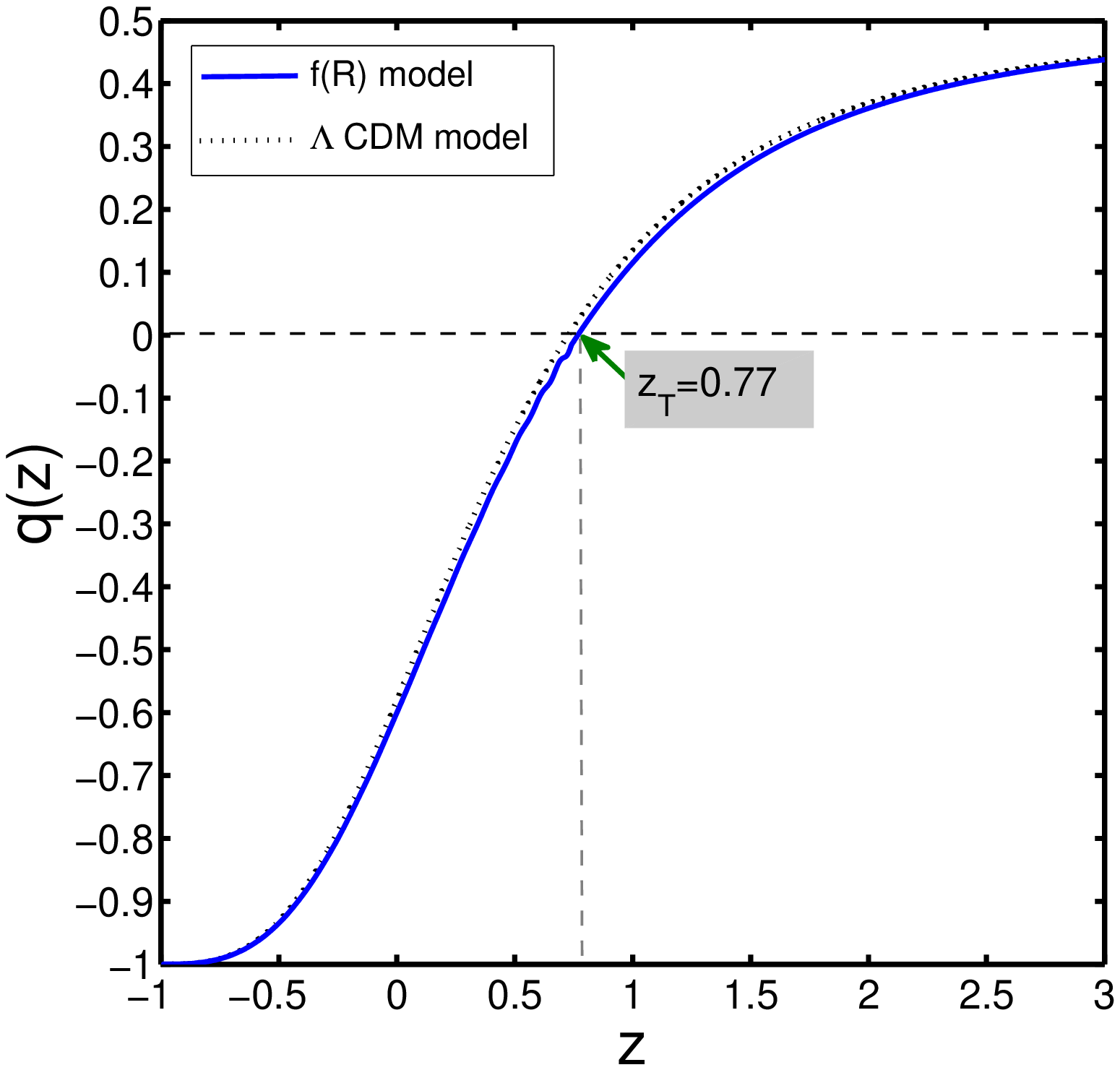}
\includegraphics[width=0.496\linewidth]{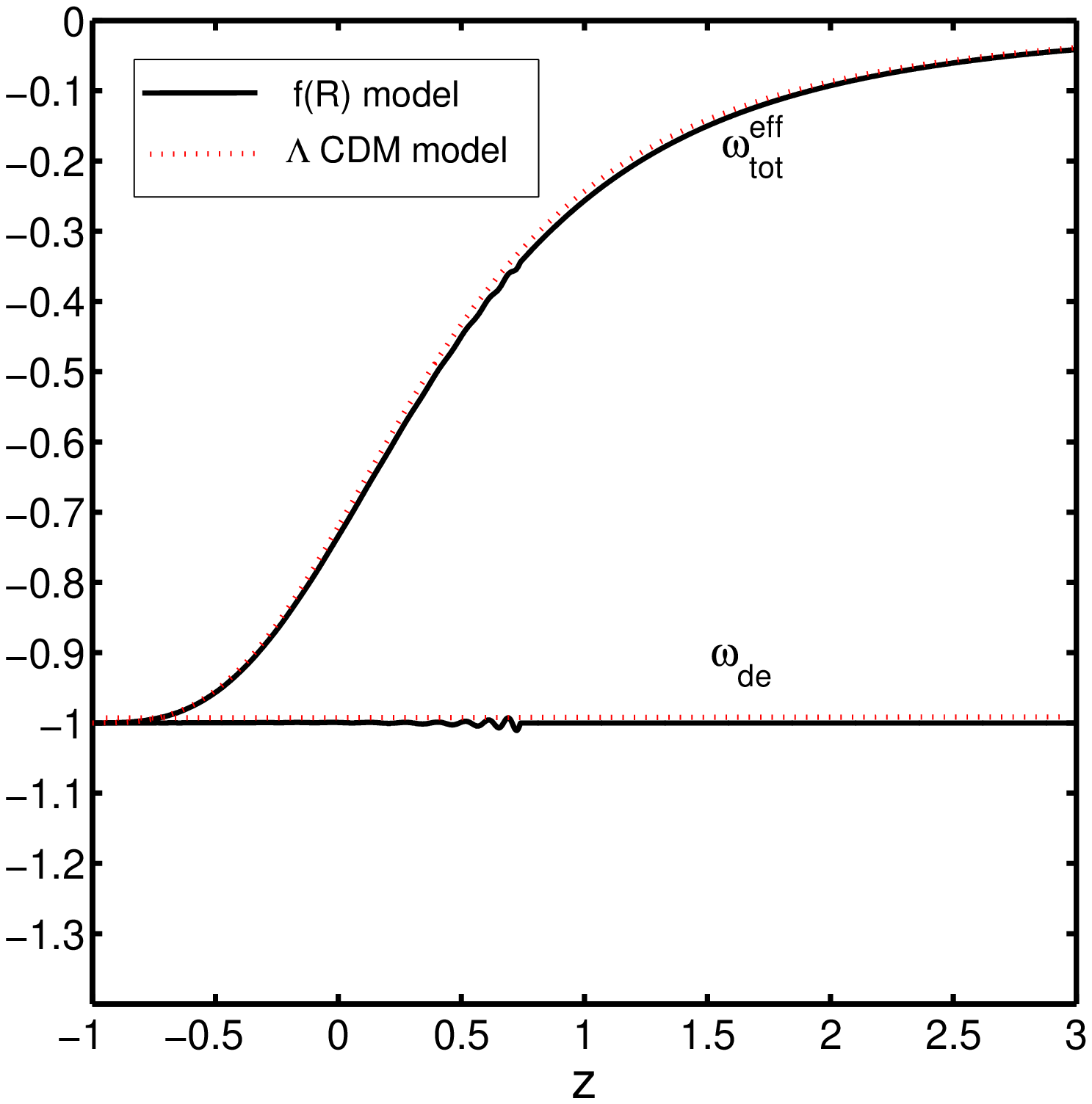}
 \caption{Evolutions of $q(z)$ (\textbf{left}), $\omega_{de}^{eff}$ and $\omega_{tot}^{eff}$ (\textbf{right})
 for the flat $\Lambda$CDM and exponential $f(R)$ models with best-fit values of corresponding parameters.}
\label{fig:qz_weff}
\end{figure}

\begin{figure}[htbp]
\centering
\includegraphics[width=0.6\textwidth,height=0.6\textwidth]{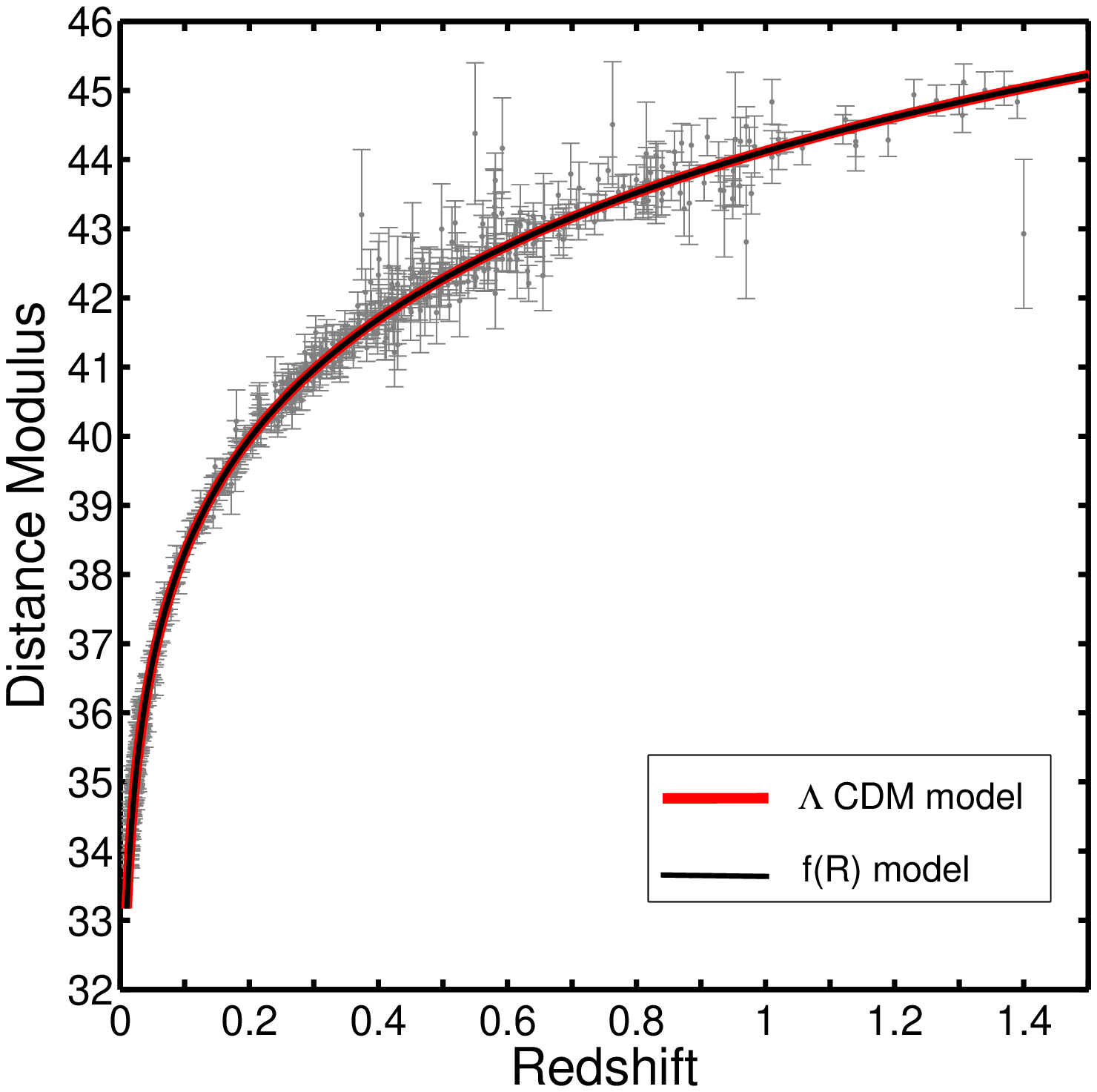}
\caption{
  Hubble diagram for the Union2 compilation of SNe Ia. The lines represent the best fitted cosmologies for the
flat $\Lambda$CDM and exponential $f(R)$ models constrained from the combination of the
 $H(z)$, BAO, CMB and SNe Ia with $\sigma_{sys}$ data discussed
in the text.}
\label{fig:mu}
\end{figure}

\begin{table*}
\caption{Results constrained from different data sets including bounds on the parameters $\Omega_m^0$ and $\beta$ at 1$\sigma$ confidence interval
 and the values of $\chi^2_{min}$, where \emph{d.o.f} denotes the degree of freedom  and $+null$ represents the absence of the upper
limit for $\beta$.}
\label{tab:results}
\begin{center}
\begin{tabular}{lccl}
\hline
Sample & $\Omega_m^0$ & $\beta$  &$\chi^2_{min}/d.o.f$ \\
\hline
$H(z)$  &$\Omega_m^0=0.235^{+0.059}_{-0.052}$  &$\beta=6.13^{+null}_{-5.13}$ &24.44/23\\
BAO  & $\Omega_m^0=0.277^{+0.043}_{-0.039}$ &$\beta=4.62^{+null}_{-3.57}$ &7.51/6 \\
$H(z)$+BAO+CMB+SNe Ia (no $\sigma_{sys}$) & $\Omega_m^0=0.267^{+0.023}_{-0.022}$ &$\beta=4.65^{+null}_{-3.11}$ &575.52/587 \\
$H(z)$+BAO+CMB+SNe Ia (with $\sigma_{sys}$) & $\Omega_m^0=0.266^{+0.030}_{-0.026}$ & $\beta=4.65^{+null}_{-3.18}$ &563.57/587  \\
\hline
\end{tabular}
\end{center}
\end{table*}
Currently, since the systematic errors in the SNe Ia data are comparable with the statistical errors,
they should be considered seriously. In our following analysis, we employ the results
from the joint analysis including the effect of SNe Ia systematic errors.
With the best-fit values of the parameters from the joint analysis of H(z), BAO, CMB
and SNe Ia with $\sigma_{sys}$, i.e., $(\Omega_m^0, \beta)=(0.266, 4.65)$, we numerically compute the evolutions of
 $q(z)$,
 $\omega_{de}^{eff}$
 and $\omega_{tot}^{eff}$,
 which are shown in Fig.~\ref{fig:qz_weff}. To make a comparison, the evolutions of
 $q(z)$, $\omega_{de}^{eff}$ and $\omega_{tot}^{eff}$ in the flat $\Lambda$CDM model with the best-fit value $\Omega_m^0=0.28$ are also displayed in Fig.~\ref{fig:qz_weff}.
 The Hubble diagram for the Union2 compilation of SNe Ia is presented in Fig.~\ref{fig:mu} with the best-fit cosmologies for the flat $\Lambda$CDM and exponential $f(R)$ models.
 It turns out that the current observational data
still cannot distinguish between the $\Lambda$CDM  and  exponential $f(R)$ models, at least for the interval of
 parameters $\Omega_m^0$ and $\beta$ given by our statistical analysis. The transition redshift $z_T$,
 at which the expansion underwent the transition from deceleration to acceleration,
 is obtained by solving the equation $q(z=z_T)=0$ or $\omega_{tot}^{eff}(z=z_T)=-1/3$.
 We work out $z_T \approx 0.77$ as marked in Fig.~\ref{fig:qz_weff}, that is in good accordance with
 the results from the literature~\cite{Cunha2008,Farooq_Ratra2013}.

\section{Conclusions}
\label{summary}

We have concentrated on the viable exponential $f(R)$ model with $f(R)=-\beta R_s(1-e^{-R/R_s})$.
In  this model, when $\beta\rightarrow \infty$, it is reduced to the corresponding $\Lambda$CDM model.
The equations of motion are derived in terms of the metric approach. We have followed the parametrization in
Refs.~\cite{Yang2010,Hu2007} to study the dynamics numerically. The current Hubble parameter measurements alone
can constrain significantly the model parameters $\Omega_m^0$ and $\beta$, while
more restrictive bounds on the parameters have been found by combining the  Hubble parameter data
with those from SNe Ia, BAO and CMB.
Explicitly, at 1$\sigma$ c.l.  for  the joint analysis of the
 $H(z)$, BAO, CMB and SNe Ia with $\sigma_{sys}$ data, we have obtained the intervals
 $0.240<\Omega_m^0<0.296$ and $1.47 <\beta< +null$. The
 range of $\Omega_m^0$ is consistent with the current observations. The absence of the upper limit for $\beta$ is due to
 that the best-fit value of $\beta$ lies in the range where the Ricci scalar $R$ is insensitive to the value of $\beta$,
 which also indicates that the background dynamics in the $\Lambda$CDM  and  exponential $f(R)$ models are nearly indistinguishable.
 To break this degeneracy, one may pin the hope on testing the evolution of perturbations in these models~\cite{DeFelice2010}.

\section*{acknowledgments}
This work was supported by the National Center for Theoretical Sciences, National
Science Council (Grant No. NSC-101-2112-M-007-006-MY3), and National Tsing-Hua University (Grant No. 103N2724E1) at Taiwan, the Ministry of Science and Technology National Basic Science Program (Project 973) under Grant
No.2012CB821804, and the National Natural Science Foundation of China.


\end{document}